\documentclass[apj,twocolumn]{emulateapj}

\slugcomment{ApJ: accepted for publication}

\shortauthors{Parmentier, Pfalzner \& Grebel}
\shorttitle{Cluster Star Age distribution}

\newcommand\eff{\epsilon_{ff}}
\newcommand\tff{\tau_{ff}}

\newcommand\mst{m_{\star}}

\newcommand\st{_{\star}}
\newcommand\Ms{M_{\odot}}

\newcommand\sfe{star formation efficiency }
\newcommand\sfr{star formation rate }

\newcommand\3{NGC~3603 }

\begin{document}



\title{Stellar age spreads in clusters as imprints of cluster-parent clump densities}


\author{G.~Parmentier\altaffilmark{1}, S.~Pfalzner\altaffilmark{2}, E.~K.~Grebel\altaffilmark{1}}


\altaffiltext{1}{Astronomisches Rechen-Institut, Zentrum f\"ur Astronomie der Universit\"at Heidelberg, M\"onchhofstr. 12-14, D-69120 Heidelberg, Germany}
\altaffiltext{2}{Max-Planck-Institut f\"ur Radioastronomie, Auf dem H\"ugel 69, 53121 Bonn, Germany }


\begin{abstract}
It has recently been suggested that high-density star clusters have stellar age distributions much narrower than that of the Orion Nebula Cluster, indicating a possible trend of narrower age distributions for denser clusters.  We show this effect to likely arise from star formation being faster in gas with a higher density.    
We model the star formation history of molecular clumps in equilibrium by associating a \sfe per free-fall time, $\eff$, to their volume density profile.  
We focus on the case of isothermal spheres and we obtain the evolution with time of their star formation rate.  Our model predicts a steady decline of the star formation rate, which we quantify with its half-life time, namely, the time needed for the \sfr to drop to half its initial value.

Given the uncertainties affecting the star formation efficiency per free-fall time, we consider two distinct values: $\eff = 0.1$ and $\eff = 0.01$.    
When $\eff = 0.1$, the half-life time is of the order of the clump free-fall time, $\tau_{ff}$. 
As a result, the age distributions of stars formed in high-density clumps have smaller full-widths at half-maximum than  those of stars formed in low-density clumps.  
When the \sfe per free-fall time is 0.01, the half-life time is 10 times longer, i.e. 10 clump free-fall times.  We explore what happens if the duration of star formation is shorter than 10$\tau_{ff}$, that is, if the half-life time of the \sfr cannot be defined.  There, we build on the invariance of the shape of the young cluster mass function to show that an anti-correlation between the clump density and the duration of star formation is expected.  We therefore conclude that, regardless of whether the duration of star formation is longer than the \sfr half-life time, denser molecular clumps yield narrower star age distributions in clusters.  Published densities and stellar age spreads of young clusters and star-forming regions actually suggest that the time-scale for star formation is of order $1$-$4\tff$.

We also discuss how the age-bin size and uncertainties in stellar ages affect our results.    
We conclude that there is no need to invoke the existence of multiple cluster formation mechanisms to explain the observed range of stellar age spreads in clusters.  
\end{abstract}



\section{Introduction}
\label{sec:intro}

A crucial constraint on models of star cluster formation is the star formation duration which reveals itself as the age spread of stars in clusters.  Cluster star age spreads are notoriously difficult to assess, as illustrated by the contrasting results reached by \citet{jef11} and \citet{pal99} for the Orion Nebula Cluster \citep[see also][for a review]{pre12}.  Nevertheless, cluster star age spreads seem to range from a fraction of a million years up to 10 million years \citep{mel08,dem11,reg11,kud12}.  Based on such a variety of results, one may conclude that different star cluster formation mechanisms exist.  Building on the concept of \sfe per free-fall time, we will show that this is not necessarily the case.  
 
The \sfe per free-fall time, first defined by \citet{kru05}, is the gas mass fraction of an object turned into stars over one free-fall time at the object mean density.  \citet{par13} recently generalized the concept by associating a {\it radially-varying} free-fall time to the observed volume density profile of proto-cluster clumps.  Note that their model applies to molecular clumps (i.e., star cluster progenitors) and not to molecular clouds (i.e., cluster complex progenitors).  In this class of models,  molecular gas with a higher volume density forms stars more rapidly than gas with a low volume density \citep[see also][]{elm00, elm11}.  
Under the assumption of a constant \sfe per free-fall time, the slope predicted for the star formation relation, namely the relation between the surface densities in molecular gas and young stars, actually matches that observed by \citet{gut11}.  This suggests that the \sfe per free-fall time is independent of the distance from the clump center, hence from the gas density.  Such a conclusion agrees with the earlier result of \citet{kru07} who found no significant variations of the \sfe per free-fall time over three orders of magnitude in gas density, from giant molecular clouds to HCN-traced molecular clumps.

Here we take the model of \citet{par13} one step further and we show that high-density molecular clumps may yield clusters with shorter stellar age spreads than those formed by low-density clumps.  Our study has been stimulated by the recent finding that stellar age spreads in the central regions of the starburst clusters NGC3603~YC and Westerlund~1 (Wd~1) may be an order of magnitude smaller than in the Orion Nebula Cluster \citep{reg11,kud12}.
In what follows, we call age spread of a cluster the full-width at half-maximum (FWHM) of its stellar age distribution, regardless of the shape of the age distribution.  
We obtain and discuss the relation between stellar age spreads in clusters and cluster-parent clump characteristics.

\section{Volume-density-dependent star-formation rates}
\label{sec:princ}
\subsection{Molecular Clump Star Formation Histories}
\label{ssec:mod}

A star-forming molecular clump has typically a power-law density profile with a slope of $-1.5$ to $-2$ \citep{mue02,pir09}.  We adopt a slope of $-2$ for the volume density profile of molecular clumps, $\rho_0(r)$, with an additional term to avoid an infinite central density:

\begin{equation}
\rho_0(r) = \rho_c \left( 1 + \frac{r}{r_c} \right)^{-2} \;.
\label{eq:cpl}
\end{equation}
$\rho_0(r)$ also defines the gas density profile at the onset of star formation. 
In this equation, $r$ is the distance to the clump center (spherical symmetry is assumed), $\rho_c$ is the central density of the clump 
and $r_c$ its core length-scale.  We take $r_c=0.01$\,pc, which is of the order of the radius of the circumstellar envelope of a protostar \citep{mot98}.  For the models discussed here, this means that $r_c \sim 0.01R$ where $R$ is the clump outer radius.  The central density $\rho _c$ is such that the mass enclosed within the clump radius $R$ is the clump total mass $M_0$: 

\begin{equation}
\rho _c = \frac{M_0(R)}{4  \pi  r_c^3 \left[  \frac{R(R+2r_c)}{r_c(R+r_c)} - 2 {\rm ln}\left( \frac{R+r_c}{r_c} \right)\right]}\;.
\label{eq:rhoc}
\end{equation}

In this model, we assume that the molecular gas first gathers into a massive and compact clump whose density profile is described by Eq.~\ref{eq:cpl}, {\it then} starts forming stars while retaining a near-equilibrium configuration.  Do such compact -- yet starless --  molecular clumps exist?  The answer seems to be yes.  In the APEX Telesope Large Area Survey of the GALaxy at 870$\mu$m \citep[ATLASGAL,][]{sch09}, \citet{tac12} identify 210 such clumps.  Their size -- of order 1\,pc -- and masses -- up to $10^4\,\Ms$ -- make them cluster-progenitor candidates.  In addition, the absence of mid-IR tracers of ongoing star formation (either Class~I young stellar objects or warm dust) lead \citet{tac12} to consider these clumps as starless.  
An alternative model is that a cluster-forming clump is contracting from volumes with radius $\gtrsim 10$\,pc over timescales of several Myr during which star formation is already proceeding.  

The key difference between both models resides in the evolution of the star formation rate.  While our model predicts a decrease with time of the star formation rate (see below), a contracting clump leads to a \sfr increase.
A third possibility is that the actual star formation history of molecular clumps consists of the combination of both models, with an increasing star formation rate during the contraction of the clump followed by a decline once stellar feedback has forced the clump to a position of near-equilibrium.  Such a scenario inevitably leads to star age distributions broader than those derived here.  

In what follows, we assume that the clump radius $R$ remains constant in time.  Such a static configuration may be maintained as protostellar outflows compensate for the decay of the turbulence inherited from the clump formation, thereby allowing the clump to resist a global gravitational contraction.  Collimated protostellar outflows are especially efficient at transporting the outflow momentum and energy to large distances, eventually driving the turbulence on the full pc-scale of the cluster-forming clump (\citet{nak07}; see \citet{mck89} for an early model).  In fact, the clump will oscillate around a position of near-equilibrium. [See also \citet{kru06} for a model of the evolution of {\it giant molecular clouds} where a cloud can oscillate around a given radius until its eventual dispersal by stellar feedback].  We will neglect such clump oscillations.

We also stress that the \sfr we consider is the stellar {\it mass} formed per time unit.  Its evolution with time may differ from the evolution of the {\it number} of stars formed per time unit, especially if high-mass stars formed first.
We come back to these points in Section~\ref{sec:acc_dec}.

Assuming that star formation starts at a time $t=0$ in an initially starless clump of mass $M_0$, radius R, and density profile given by Eq.~\ref{eq:cpl}, its \sfr (SFR) at time $t$ is:
\begin{eqnarray}
SFR(t) & = & \int_0^R \frac{\partial \rho \st (t,r)}{\partial t}  4 \pi r^2 {\rm d} r\;.
\label{eq:sfr1}
\end{eqnarray}
Here, $t$ is the time elapsed since the onset of star formation, $\rho \st (t,r)$ is the radial density profile of the stars at time $t$ and $\partial \rho \st (t,r)/\partial t$ describes therefore the evolution with time of the volume density of the star formation rate. 

Equation~\ref{eq:sfr1} can be written as a function of the initial gas density profile, $\rho_0(r)$, and \sfe per free-fall time, $\epsilon_{ff}$, by combining Eqs.~18-19 of \citet{par13}: 
\begin{eqnarray}
SFR(t) =     \qquad    \qquad     \qquad           \qquad    \qquad     \qquad  \qquad    \qquad     \qquad  \nonumber & \\ 
\int_0^R \sqrt{\frac{32G}{3\pi}} \eff \left( \rho_0(r)^{-1/2} + \sqrt{\frac{8G}{3\pi}} \eff t \right)^{-3} 4 \pi r^2 {\rm d} r \;. &
\label{eq:sfr2}
\end{eqnarray}

One model parameter which still needs to be specified is the \sfe per free-fall time, $\eff$.  Based on the relation between the observed surface densities of the gas and star formation rate in unresolved disk and starburst galaxies in the local and high redshift Universe, as well as in kpc-scale regions of Local Group galaxies, \citet{kru12} infer $\eff \simeq 0.01$ with $0.003 \lesssim \eff \lesssim 0.03$.  Their result is similar to that obtained by \citet{kru07}, from CO-traced clouds to HCN-traced clumps (i.e. clumps with a mean density of about $n_{H2} \simeq 10^4\,cm^{-3}$).
In contrast, \citet{par13} derive an-order-of-magnitude higher estimate, i.e. $\eff \simeq 0.1$, by matching their model onto the observed star formation relation of \citet{gut11}: 

\begin{equation}
\frac{\Sigma_{\star}}{M_{\odot}\cdot pc^{-2}} = 10^{-3} \left(\frac{\Sigma_{g}}{M_{\odot}\cdot pc^{-2}}\right)^2\;.
\label{eq:sfl}
\end{equation}

\begin{figure}
\begin{center}
\epsscale{1.}  \plotone{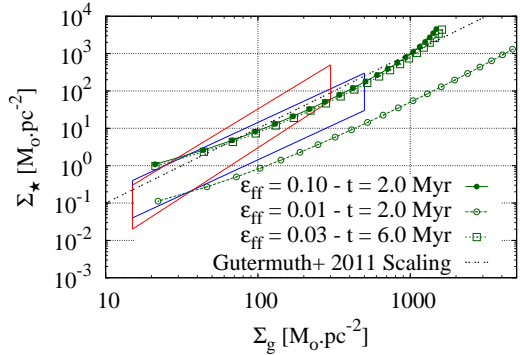}
\caption{
Local surface density of YSOs, $\Sigma_{\star}$, in dependence on the local surface density of the unprocessed/observed gas, $\Sigma_{g}$.  The dotted (black) line depicts the averaged star formation relation inferred by \citet{gut11} for a sample of molecular clouds of the Solar neighborhood and given by Eq.~\ref{eq:sfl}.  The red and blue polygons illustrate the observations for the MonR2 and Ophiuchus molecular clouds, where the scatter is the smallest.  These observational results are compared with three models of \citet{par13}, for a clump with a mass $M_0 = 10^4\,\Ms$, a radius $R = 6\,$pc, and a density profile described by Eq.~\ref{eq:cpl}.  The time intervals since the onset of star formation and the star formation efficiencies per free-fall time are given in the key.  Additional information about this model can be found in \citet{par13} \label{fig:eff}
}
\end{center} 
\end{figure}

In Eq.~\ref{eq:sfl}, $\Sigma_{g}$ and $\Sigma_{\star}$ are the local surface densities of gas and young stellar objects (YSOs) in molecular clouds of the Solar Neighborhood.  Figure \ref{fig:eff} shows Eq.~\ref{eq:sfl} (black dotted line) as well as two models of \citet{par13}: $\epsilon_{ff}=0.1$ and $t=2$\,Myr (filled circles), and $\epsilon_{ff}=0.01$ and $t=2$\,Myr (open circles).  Although the model with $\epsilon_{ff}=0.1$ provides a better match to Eq.~\ref{eq:sfl} than its $\epsilon_{ff}=0.01$ counterpart, one should keep in mind that the $\eff$ estimate is directly related to the assumed time-span $t$ since the onset of star formation.  For a given \sfe per free-fall time, the longer the time-span, the higher the surface density in YSOs.  \citet{par13} assume an averaged star-formation timespan of $t=2$Myr in the star-forming regions studied by \citet{gut11}.  This estimate is based on the presence of Class~II objects in these regions and on an estimated duration of the Class~II phase of $t_{II} \simeq 2$\,Myr \citep{eva09}.
Should the Class~II phase be longer, the estimated \sfe per free-fall time would be smaller (since a smaller $\eff$ is needed to achieve the same stellar surface densities over a longer star-formation time-span).  

The assumption of a 2Myr time-scale for the Class II phase 
has recently been challenged by \citet{bell13}.  They study star-forming regions where massive stars have already settled on the main sequence and infer a Class~II phase duration of $t_{II} \simeq 4-5$\,Myr.  This is about twice as long as the earlier result of \citet{eva09}.  It also implies a longer Class~I phase duration, i.e., $t_I \simeq 1$\,Myr.  If the result of \citet{bell13} stands, and if star formation in the regions surveyed by \citet{gut11} covers the full Class~II phase, then the time-span since the onset of star formation is closer to $t \simeq t_I + t_{II} \simeq 5-6$\,Myr, a factor of 3 longer than assumed by \citet{par13}.  That would accordingly decrease their $\eff$ estimate by a factor 3 ($\eff \simeq 0.03$), bringing it in agreement with the upper limit on the result of \citet{kru12}.  This degeneracy between the time elapsed since star formation onset and the \sfe per free-fall time is illustrated in Fig.~\ref{fig:eff}.  One can see that the models with $t=6$\,Myr and $\eff=0.03$ on the one hand, and with $t=2$\,Myr and $\eff=0.10$ on the other hand, lead to identical star formation relations. 
 
\begin{figure}
\begin{center}
\epsscale{1.}  \plotone{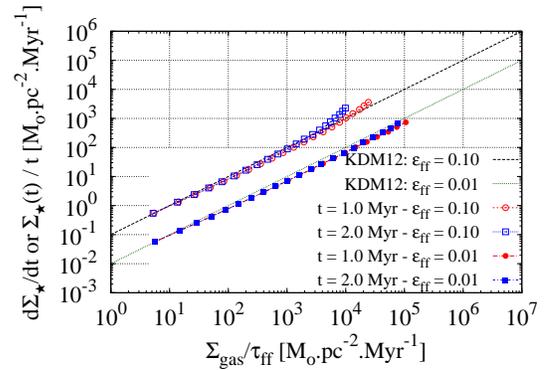}
\caption{
Star formation relation as introduced by \citet{kru12} [KDM12].  The symbol-free lines depict their eq.~2 (Eq.~\ref{eq:kdm12} in this paper).  The lines with squares and circles illustrate the predictions of the model of \citet{par13} for the same clump mass and radius as in Fig.~\ref{fig:eff}.   
\label{fig:eff_kdm}
}
\end{center} 
\end{figure}

That \citet{kru12} and \citet{par13} derive different estimates of the \sfe per free-fall time might also be related to the use of  different star formation relations.  While \citet{par13} build on the $\Sigma_{\star}$ vs. $\Sigma_{gas}$ space of \citet{gut11}, \citet{kru12} work in the $\dot{\Sigma}_{\star}$ vs. $\Sigma_{gas}/\tff$ space.  Specifically, they obtain the $\eff$ by matching their collected data onto their eq.~2 which we reproduce here for the sake of clarity:

\begin{equation}
\dot{\Sigma}_{\star} = f_{H2} \eff \frac{\Sigma_{gas}}{\tff}\;.
\label{eq:kdm12}
\end{equation}

$f_{H_2}$ is the fraction of molecular gas which we take equal to unity.  Note that the slope of Eq.~\ref{eq:sfl} is two, while the slope of the star formation relation implemented by \citet{kru12} (Eq.~\ref{eq:kdm12}) is unity.  To test whether this can lead to different $\eff$ estimates, we plot in Fig.~\ref{fig:eff_kdm} Eq.~\ref{eq:kdm12} for $\eff = 0.01$ (green dotted line) and $\eff = 0.10$ (black dashed line).  The green dotted line ($\eff = 0.01$) is the star formation relation shown in fig.~3 of \citet{kru12}.  In the same figure, we plot the outputs of the \citet{par13} model for the same efficiencies per free-fall time and two different times after the onset of star formation ($t=1$\,Myr and $t=2$\,Myr).  We approximate $\dot{\Sigma}_{\star}$ by ${\Sigma}_{\star}(t)/t$. 
$\tau_{ff}$ is the free-fall time relevant for star formation, namely, this is the free-fall time of the gas in a spherical shell whose radius is the distance from the clump center at which $\Sigma_{\star}$ and $\Sigma_{gas}$ are measured.  In other words, $\tau_{ff}$ is a three-dimensional, local and instantaneous measurement ($\tff = \tff(r,t)$).  For a given \sfe per free-fall time, Fig.~\ref{fig:eff_kdm} does not reveal any significant difference between the star formation relations of \citet{kru12} (dashed and dotted lines) and the predictions of our model (lines with symbols).  We therefore conclude that the difference between the $\eff$ estimates of \citet{kru12} and \citet{par13} does not stem from using two different star formation relations.

Given the uncertainties and caveats described above, in what follows, we apply Eq.~\ref{eq:sfr2} for two values of the \sfe per free-fall time, i.e. $\eff \simeq 0.01$ and $\eff \simeq 0.10$.  $\eff$ is assumed to be constant in time and uniform in space.  As we shall see, these two values lead to two different regimes when defining the stellar age spread of a cluster.

We now proceed to obtain the evolution with time of the star formation rate.  Combining Eq.~\ref{eq:cpl} and Eq.~\ref{eq:sfr2}, one obtains:
\begin{eqnarray}
SFR(t) & = & 8 \pi \sqrt{\frac{8G}{3\pi}} \cdot \eff \cdot \rho_c^{3/2} r_c^3 \nonumber \\ 
       & \times & \left[ {\rm ln}\left( \frac{R+\mathcal{C}r_c}{\mathcal{C}r_c} \right) - \frac{3}{2} + \frac{\mathcal{C}r_c ( 4R + 3\mathcal{C}r_c)}{2(R+\mathcal{C}r_c)^2} \right] \,,
\label{eq:sfrcpl}
\end{eqnarray}

where $\mathcal{C}$ is a function of time $t$:
\begin{equation}
\mathcal{C} = 1 + \sqrt{\frac{8G}{3\pi}} \cdot \eff \cdot \rho_c^{1/2} \cdot t \;.
\label{eq:cround}
\end{equation}

\begin{figure}
\begin{center}
\epsscale{1.}  \plotone{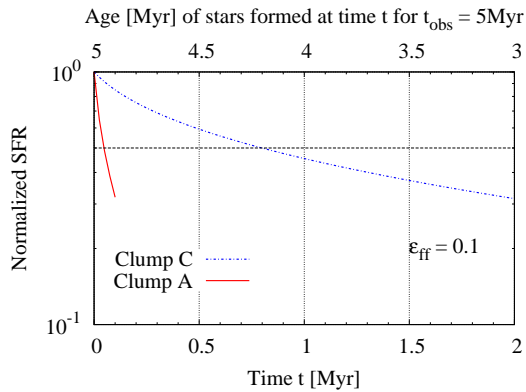}
\caption{
Evolution with time of the \sfr normalized to its initial value for two molecular clumps (gas initial density profile given by Eq.~\ref{eq:cpl} and characteristics listed in Table \ref{tbl:t1}) when $\eff = 0.1$.
The bottom $x$-axis gives the time elapsed since the onset of star formation.  The duration of star formation is $5\tau_{ff}$, equivalent to 0.13Myr and 3.5Myr for Clump~A and Clump~C, respectively.  As for Clump~C, note that the figure is limited to the first 2\,Myr of evolution.  The top $x$-axis corresponds to the age of the newly-formed stars if the cluster is observed 5\,Myr after star formation onset.  The horizontal thin dashed line indicates half the initial \sfr   
}
\label{fig:princ}
\end{center} 
\end{figure}

\begin{figure}
\begin{center}
\epsscale{1.}  \plotone{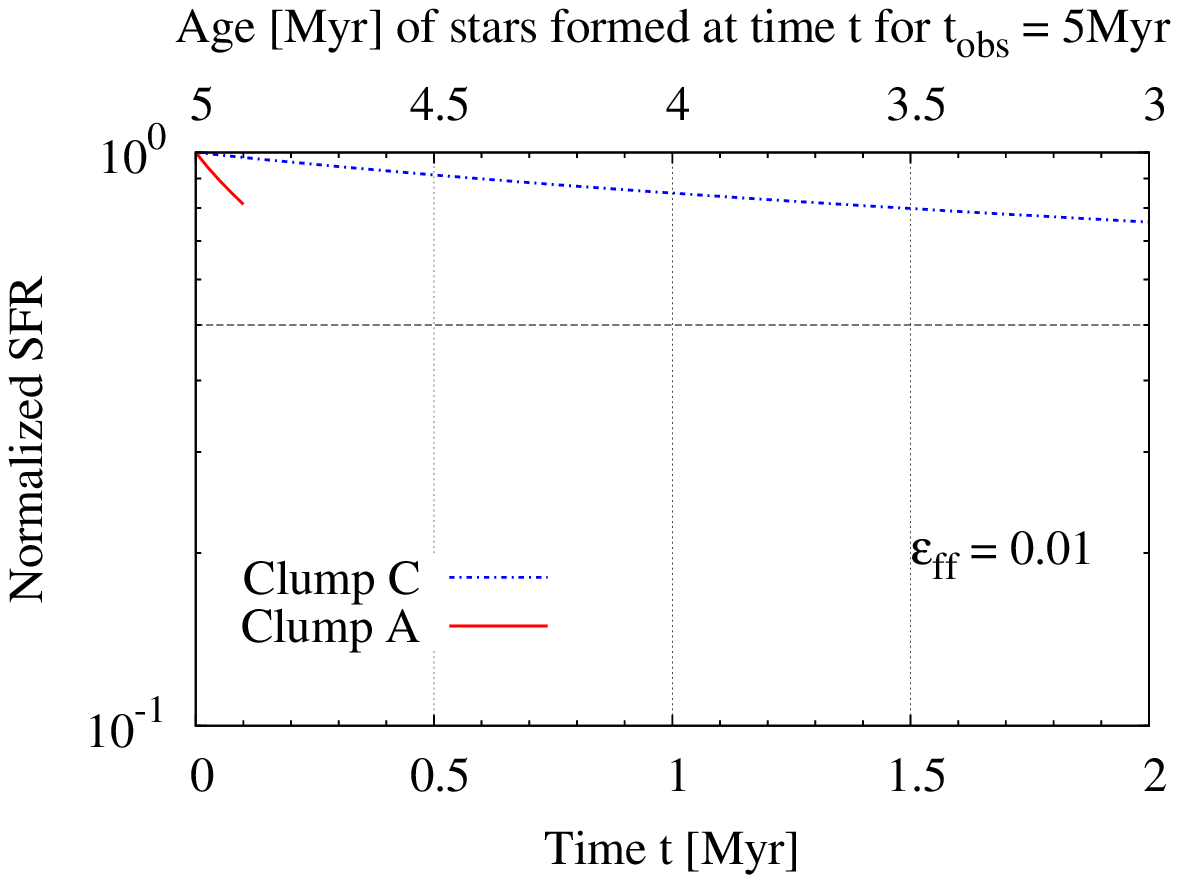}
\caption{
Same as Fig.~\ref{fig:princ}, but for a \sfe per free-fall time of $\eff = 0.01$. 
}
\label{fig:princb}
\end{center} 
\end{figure}

Figures~\ref{fig:princ} and \ref{fig:princb} present the evolution of the SFR normalized to its initial value for two molecular clumps, Clump~A and Clump~C, and two star formation efficiencies per free-fall time, $\eff = 0.1$ (Fig.~\ref{fig:princ}), and $\eff = 0.01$ (Fig.~\ref{fig:princb}).  The characteristics of the two model clumps are listed in Table \ref{tbl:t1}.  Table \ref{tbl:t1} also gives the mean free-fall time of the gas at star formation onset:
\begin{equation}
\tau _{ff}(t=0) = \sqrt{ \frac{3\pi}{32G\langle\rho_0\rangle} }
\label{eq:tff}
\end{equation}
where $\langle\rho _0\rangle$ is the clump mean density.  

The mass and radius of Clump~C are similar to what was found by \citet{hil98} for the gas-embedded Orion Nebula Cluster (ONC).  Clump~C ($\langle\rho_0\rangle\simeq 100\,\Ms.pc^{-3}$) may thus be a good proxy to the ONC parent clump.  
 
Given our assumption that the clump remains in equilibrium during star formation, we set an upper limit of five initial free-fall times to the duration of star formation.  This represents 0.13\,Myr and 3.5\,Myr for Clump~A and Clump~C, respectively (see Table \ref{tbl:t1}).  During this time-span, the clump gas content is solely depleted by the conversion of gas into stars.  That is, we ignore the possibility that all or a fraction of the gas may be expelled beyond the clump radius $R$ by stellar feedback processes.  Should gas removal take place before five initial free-fall times, star formation would be slowed down through a decrease of the gas mass and an expansion of the gas clump.    

As star formation depletes the clump gas reservoir, the gas free-fall time steadily increases and the \sfe per free-fall time is applied to an ever lower gas mass.  Therefore, unless the clump contracts or is replenished in gas by inflows, the \sfr decreases with time. This decrease is faster at higher $\eff$, as well as in high-density regions (here Clump~A) since -- in essence -- they evolve on a faster time-scale.  

\begin{table}[t]
\begin{center}
\caption{Model clump characteristics.  $<\Sigma_0>$ is the clump mean surface density, $n_{H_2}$ its mean volume number density; see text for the meaning of the other symbols}  
\label{tbl:t1}
\begin{tabular}{lccc}
\tableline\tableline
                                  &  ~~Clump A~~          &   ~~Clump C~~         \\ \tableline
R [pc]                            &    0.5                & 2.0                 \\
$M_0 [\Ms]$                       & $6 \cdot 10^4$        & $4 \cdot 10^3$ \\
$\langle \rho_0\rangle [\Ms \cdot pc^{-3}]$    & $1.2 \cdot 10^5$      &  $1.2 \cdot 10^2$ \\
$\langle \rho_0\rangle [g \cdot cm^{-3}]$      & $7.8 \cdot 10^{-18}$  &  $7.8 \cdot 10^{-21}$ \\
$n_{H_2} [cm^{-3}]$               & $1.6 \cdot 10^6$      &  $1.7 \cdot 10^3$ \\
$<\Sigma_0> [\Ms \cdot pc^{-2}]$  & $7.6 \cdot 10^4$      &  $3.2 \cdot 10^2$ \\
$<\Sigma_0> [g \cdot cm^{-2}]$    & $1.6 \cdot 10^{1}$  &  $6.7 \cdot 10^{-2}$ \\
$\tff (t=0)$ [Myr]                & $0.02$                & 0.74             \\ 
$r_c$ [pc]                       & 0.01              &  0.01             \\
$\rho_c [\Ms \cdot pc^{-3}]$     & $1.1 \cdot 10^8$  &  $1.7 \cdot 10^6$ \\ 
$t_{1/2}$ [Myr] ($\epsilon_{ff}=0.10$)                 & 0.05              &  0.80     \\ \tableline
\end{tabular}
\end{center}
\end{table}

Given that the model is valid for a few free-fall times and that $r_c << R$, Eq.~\ref{eq:sfrcpl} can be simplified.  Introducing $\rho_e = \rho(r=R)$, the gas volume density at the edge of the clump, and using Eq.~\ref{eq:cpl}, we find:

\begin{equation}
\rho_e \cdot R^2 = \rho_c \cdot r_c^2\;.
\label{eq:coredge}
\end{equation}

One can then express the product $\mathcal{C} r_c$ as a function of $\tau_{ff,e}$, the free-fall time at the clump edge.  From Eq.~\ref{eq:cround}, we successively derive:
\begin{eqnarray}
\mathcal{C} r_c & = & r_c + \sqrt{\frac{8G}{3\pi}} \eff \rho_c^{1/2} r_c t \\ \nonumber
                & = & r_c + \sqrt{\frac{8G}{3\pi}} \eff \rho_e^{1/2} R t  \\ \nonumber
                & = & r_c + \frac{\eff}{2} \frac{t}{\tau_{ff,e}} R \;.
\label{eq:simpl1}
\end{eqnarray}

Given the limitations imposed to the parameter space, that is, $r_c << R$, $t \lesssim$ several $\tau_{ff,e}$ and $\eff \leq 0.1$, one gets:

\begin{equation}
R + \mathcal{C} r_c \simeq R\,,
\label{eq:simpl2}
\end{equation}

and Eq.~\ref{eq:sfrcpl} becomes:  

\begin{eqnarray}
SFR(t) & \simeq & 8 \pi \sqrt{\frac{8G}{3\pi}} \cdot \eff \cdot \rho_c^{3/2} r_c^3 \\ \nonumber
       & \cdot  & \left[ - ln \left( \frac{r_c}{R} + \frac{\eff}{2} \frac{t}{\tau_{ff,e}} \right) - \frac{3}{2} + 2\frac{\mathcal{C} r_c}{R}\right] \\ \nonumber
       & \simeq & 4 \pi \sqrt{\frac{32G}{3\pi}} \eff \rho_e^{3/2} R^3 \\ \nonumber
       & \cdot  & \left[ - ln \left( \frac{r_c}{R} + \frac{\eff}{2} \frac{t}{\tau_{ff,e}} \right) - \frac{3}{2} + 2\frac{r_c}{R} + \eff \frac{t}{\tau_{ff,e}}\right]\;. \\ \nonumber
       & \simeq & \frac{4\pi \eff \rho_e R^3}{\tau_{ff,e}} \\ \nonumber
       & \cdot  & \left[ - ln \left( \frac{r_c}{R} + \frac{\eff}{2} \frac{t}{\tau_{ff,e}} \right) - \frac{3}{2} + \eff \frac{t}{\tau_{ff,e}}\right]\;.
\end{eqnarray}

The mean density of an isothermal sphere is three times the density at its edge, i.e. $\langle \rho_0 \rangle = 3\rho_e$.  Therefore, $\tau_{ff,e} = \sqrt{3} \tau_{ff}$, with $\tau_{ff}$ the free-fall time at the clump mean density, and:
\begin{equation}
SFR(t) = \eff \frac{M_0}{\sqrt{3} \tau_{ff}} 
\left[ - ln \left( \frac{r_c}{R} + \frac{\eff}{2} \frac{t}{\sqrt{3} \tau_{ff}} \right) - \frac{3}{2} + \eff \frac{t}{\sqrt{3} \tau_{ff}}\right]\;.
\label{eq:sfr_simpl}
\end{equation}
Equation \ref{eq:sfr_simpl} is shown as the dashed green lines in the top panels of Figs.~\ref{fig:sfh} and \ref{fig:sfhb} which we discuss in the next section.
It can be integrated to provide the growth with time of the global star formation efficiency of the clump.  With $\chi = r_c/R$ and $\xi =(\eff t )/(\sqrt{3} \tau_{ff})$, we derive:
\begin{equation}
SFE(t) = \frac{M_*(t)}{M_0} = 2\chi ln(\chi) -2\left( \chi + \frac{\xi}{2}\right) \cdot ln\left( \chi + \frac{\xi}{2}\right) - \frac{\xi}{2} + \frac{1}{2} \xi^2\;.
\label{eq:sfet}
\end{equation}
We emphasize that this equation is valid only when $r_c << R$, $\eff \leq 0.1$ and for star formation durations not longer than several free-fall times.

\subsection{From formation time to star age}
\label{ssec:modobs}

To derive the stellar age distribution, we need the stellar mass formed at time $t$ over a constant time-interval $\Delta t$.  Let us assume that $\Delta t$ is small enough so that the \sfr does not vary significantly over an interval $[t,t+\Delta t]$.  The stellar mass formed over such a time interval is then:
\begin{equation}
\mst(t, \Delta t) = SFR(t) \cdot \Delta t\;.
\label{eq:mstt_cpl}
\end{equation}
If the cluster is observed at a time $t_{obs}$ after star formation onset, its stellar age distribution is:
\begin{equation}
\mst(t_{obs}-t, \Delta t) = SFR(t) \cdot \Delta t
\label{eq:mstt_cpl2}
\end{equation}
where $t_{obs}-t$ is the age of the stars formed at time $t$.  In other words, for constant and small enough $\Delta t$, the function $SFR(t)$ quantifies the stellar age distribution once time $t$ is swapped for the star age $t_{obs}-t$.  
The top $x$-axis of Figs~\ref{fig:princ} and \ref{fig:princb} shows the age of the stars formed at time $t$ for $t_{obs}=5$\,Myr.  
Therefore, this top $x$-axis provides us with the normalized star age distribution of the newly-born cluster.   Here, we assume that no stars are lost from the cluster between the time of their formation and the time of cluster observation.  Reasons for a cluster to lose stars at an early stage of its evolution include star-encounter-driven scattering \citep{pfa13} and gas expulsion \citep{par12}.

Here, a caveat regarding the size of the $\Delta t$ interval (equivalent to the bin size of the age distribution).  It cannot necessarily be chosen as small as requested by the \sfr variations.  For instance, the age bin must be large enough so as to contain enough stars to limit the amplitude of the Poisson noise.  There may thus be cases where the \sfr varies so rapidly that it cannot be considered constant over the adopted time interval $\Delta t$.  In other words, the stellar age distribution of a cluster cannot necessarily capture all the star formation rate variations of its gaseous progenitor.  This happens preferentially for high-density clumps and/or high \sfe per free-fall time.  The next section will present such a case.

\begin{figure}
\begin{center}
\epsscale{1.0}  \plotone{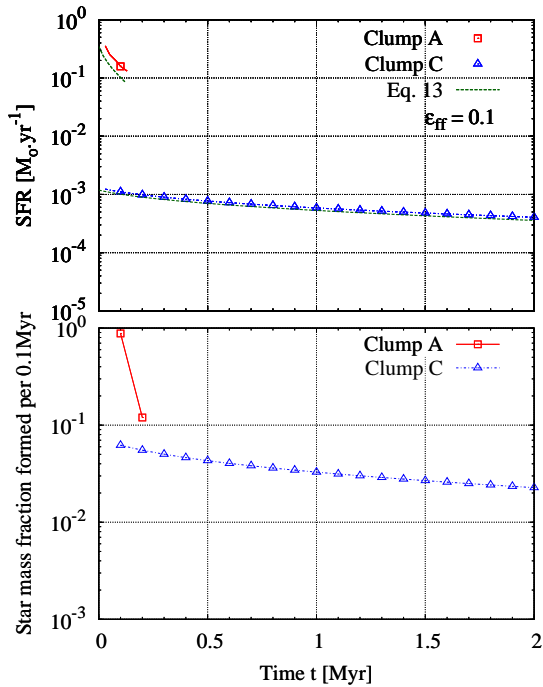}
\caption{{\it Top panel:} Evolution with time of the \sfr of the molecular clumps of Fig.~\ref{fig:princ}.  Lines with symbols depict the exact solution (Eq.~\ref{eq:sfrcpl}).  The (green) dashed symbol-free lines represent the approximation given by Eq.~\ref{eq:sfr_simpl}.  {\it Bottom panel:} Evolution with time of the star mass fraction formed per 0.1Myr.  The mass fraction is defined with respect to the total stellar mass formed $5\tff$ after the onset of star formation.  This represents 0.13Myr and 3.5\,Myr for Clump~A and Clump~C, respectively.  As for Clump~C, note that the figure is limited to the first 2\,Myr of evolution.}
\label{fig:sfh}
\end{center} 
\end{figure}

If one wants to define the cluster star age spread as the FWHM of the stellar age distribution, we see two possible regimes depending on how fast the \sfr decreases with time.  We introduce them here, and present the corresponding age distributions in the next section.  
\begin{itemize}
\item [\it (i)] If the \sfr decreases by more than a factor of 2 over the duration of star formation (here when $\eff = 0.10$; see Fig.~\ref{fig:princ}), the FWHM of the age distribution mirrors the time $t_{1/2}$ required for the \sfr to drop by a factor of 2 compared to its initial value.  We refer to $t_{1/2}$ as the half-life time of the clump star formation rate.  Half the initial \sfr is visualized in Figs~\ref{fig:princ}  and \ref{fig:princb} by the horizontal (black) dashed line, and the $t_{1/2}$ time of each model clump is given in Table \ref{tbl:t1}.  We see that $t_{1/2} \simeq \tau_{ff}\propto (\langle\rho_0\rangle)^{-1/2}$, which suggests that, in the $\eff = 0.10$ regime, the star age distribution of a young cluster can be used to probe its parent-clump volume density.  Note that $t_{1/2}$ also depends on the gas initial density profile, with shallower profiles yielding longer half-life times.  For instance, a top-hat profile implies $t_{1/2} \simeq 5 \tau _{ff}(t=0)$ when $\eff = 0.10$.  
\item [\it (ii)] If the decrease of the \sfr is so slow that $t_{1/2}$ cannot be defined (here when $\eff = 0.01$, see Fig.~\ref{fig:princb}), the FWHM of the age distribution is primarily determined by the duration of star formation.
\end{itemize}

\begin{figure}
\begin{center}
\epsscale{1.0}  \plotone{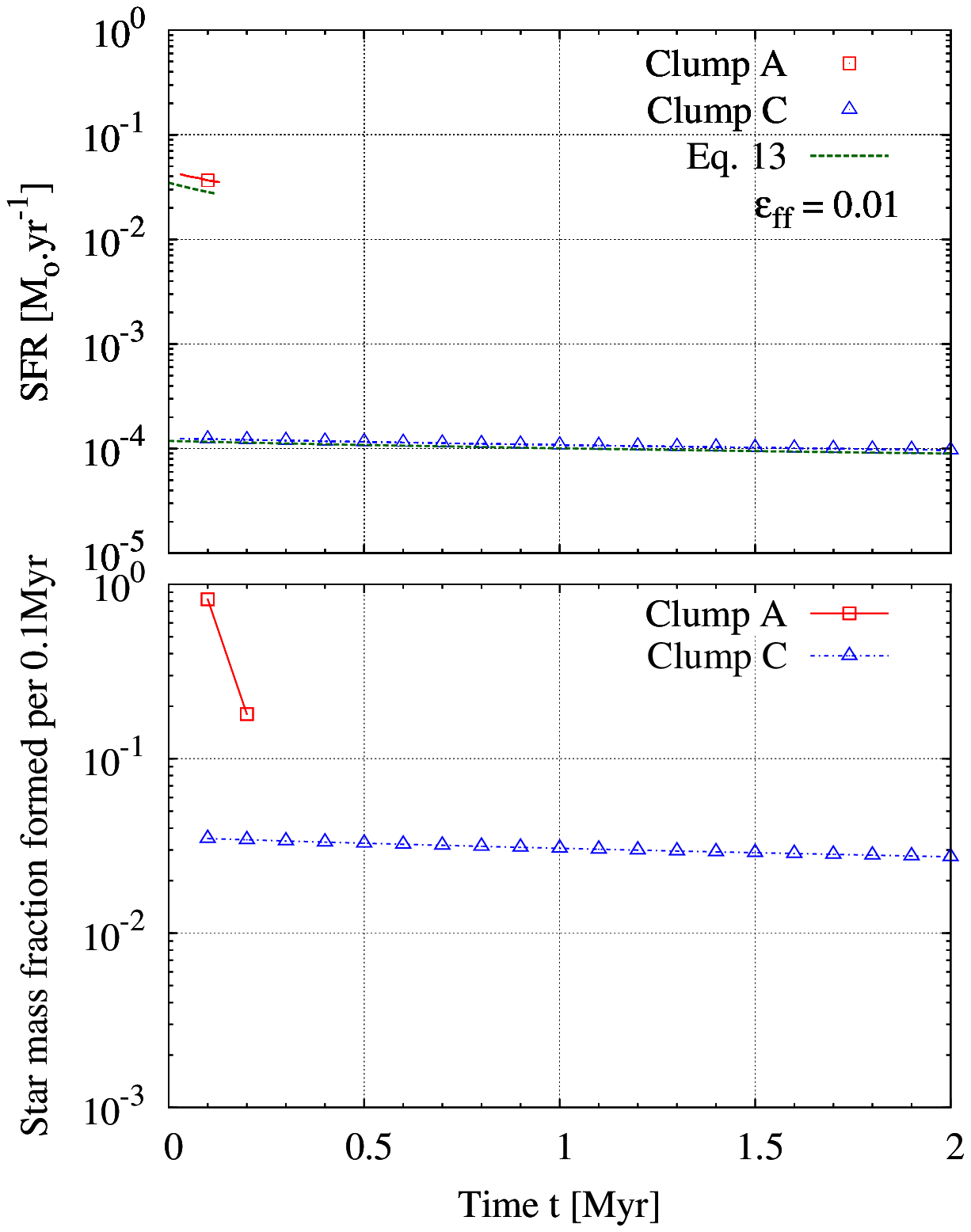}
\caption{
Same as Fig.~\ref{fig:sfh} but for $\eff = 0.01$.
}
\label{fig:sfhb}
\end{center} 
\end{figure}

The top panels of Figs~\ref{fig:sfh} and \ref{fig:sfhb} show the evolution of the star formation rate without the normalization of Figs~\ref{fig:princ} and \ref{fig:princb}.  

The bottom panels depict the time evolution of the stellar mass fraction formed per constant time-interval $\Delta t$.  For the sake of an easy comparison, we choose $\Delta t=0.1$\,Myr, identical to the age-bin width of the stellar age distributions presented in the next section.  
The mass fractions are defined with respect to the final stellar mass, here the clump stellar mass at  $t_{end}=5\tff$ (3.5\,Myr for Clump~C and 0.13\,Myr for Clump~A).  As an example, when $\eff = 0.01$, Clump~C forms at any time $\simeq 3$\,\% of its final stellar mass per time-interval of 0.1\,Myr (Fig.~\ref{fig:sfhb}).
Note also that the \sfr is approximately constant when $\eff = 0.01$ since only a few per cent of the gas are turned into stars.

\section{Age distributions within newly-born star clusters}
\label{sec:ad}

\begin{figure}
\begin{center}
\epsscale{1.0}  \plotone{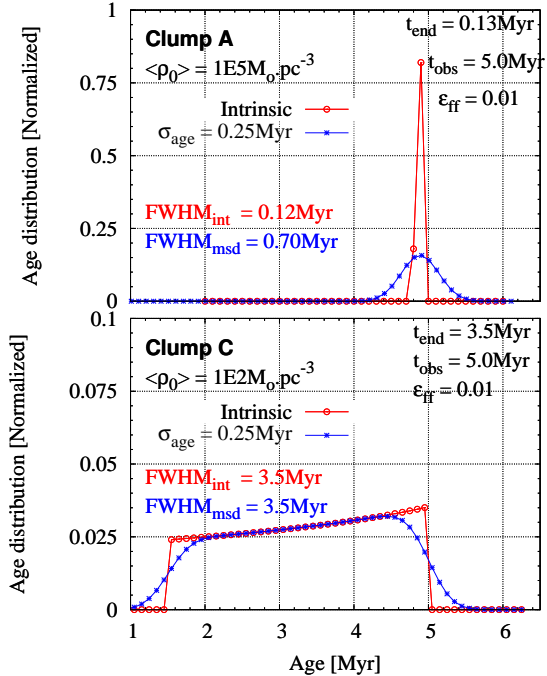}
\caption{Star age distributions for the star formation histories of Figs.~\ref{fig:princ}-\ref{fig:sfhb}.  Intrinsic and measured age distributions are shown as (red) lines with open circles and (blue) lines with asterisks.  
Note the decrease of the $y$-axis extent from the top (Model~A) to the bottom panel (Model~C).}
\label{fig:ad}
\end{center} 
\end{figure}

\begin{figure}
\begin{center}
\epsscale{1.0}  \plotone{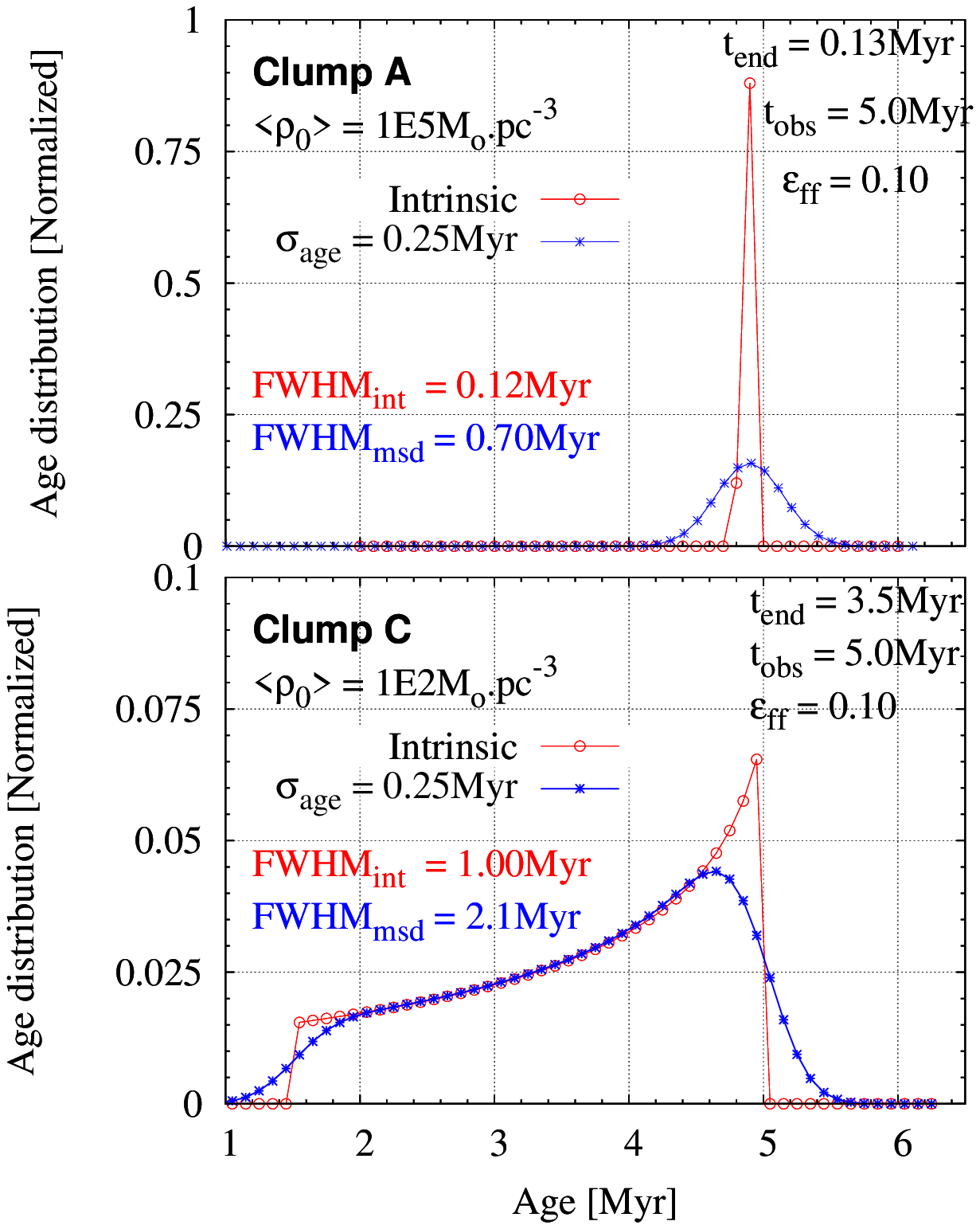}
\caption{Same as for Fig.~\ref{fig:ad} but with $\eff = 0.1$.}
\label{fig:adb}
\end{center} 
\end{figure}

We now move to estimating the star age distributions generated by our model clumps.  They are presented in  
Figs~\ref{fig:ad} ($\eff = 0.01$) and \ref{fig:adb} ($\eff = 0.1$).  The solid red lines with open circles are the intrinsic (i.e. error-free) age distributions.  Star formation started 5\,Myr ago and lasted for 0.13\,Myr and 3.5\,Myr for Clump~A and Clump~C, respectively.  Each age distribution is normalized with respect to the total stellar mass eventually formed.  
For the sake of an easy comparison with their results, we adopt the same age-bin size as \citet{kud12}, that is, 0.1\,Myr.  We come back to this choice later in this section.
These intrinsic distributions thus correspond to the models shown in the bottom panels of Figs~\ref{fig:sfh} and \ref{fig:sfhb} mirrored around a vertical axis, that is, the time $t$ since star formation onset has been swapped for the age of the stars formed at time $t$.  Note that the $y$-axis scaling is logarithmic in Figs~\ref{fig:sfh} and \ref{fig:sfhb}, and linear in Figs~\ref{fig:ad} and \ref{fig:adb}.  

The intrinsic age spreads, $FWHM_{int}$, are quoted in Figs~\ref{fig:ad} and \ref{fig:adb}.  Clump~C clearly illustrates the two regimes introduced in the previous section.

In the first regime, when $\eff = 0.01$ (bottom panel of Fig.~\ref{fig:ad}), the \sfr does not decrease significantly over the duration of the star-formation episode.  As a result, its half-life time cannot be defined, and the FWHM of the intrinsic age distribution is driven by the duration of star formation, $\Delta t_{SF}$. 
This happens if the gas density is low and/or $\eff$ is small.  A constraint on the duration of star formation as a function of the cluster-forming gas density is provided by the study of cluster mass functions.  
The mass function of young clusters has a power-law of slope $\simeq -2$ , i.e. $dN \propto m^{-2} dm$.  At young ages (e.g., age $<$ 50\,Myr), it retains its shape while evolving towards lower masses \citep{lad03, cha10}.  This means that, following the expulsion of their residual star-forming gas, all clusters experience the same mass fraction of infant weight-loss, regardless of their mass.  In turn, this implies that all cluster parent-clumps achieve, irrespective of their mass, similar star formation efficiencies at the time of gas expulsion.  For a given star formation efficiency per free-fall time, a common \sfe suggests a star formation duration equal to a common number $k$ of free-fall times, namely, $\Delta t_{SF} = k \cdot \tau_{ff}$ for all cluster-forming clumps (see Eq.~\ref{eq:sfet}).  
The invariance of the cluster mass function shape at young ages therefore suggests a star formation duration in clusters scaling with the clump free-fall time and, hence, shorter age spreads at higher clump densities.  

Note, however, that the formation of a bound gas-free star cluster in $5 \tau_{ff}$ with $\eff = 0.01$ requires an extremely large amount of gas.  With a final \sfe of $\simeq$5\,\%, the star mass bound fraction is about 10\,\% \citep[see fig.~15 in][]{par13}.  That is, the formation of a bound gas-free cluster with a mere mass of $10^3\Ms$ requires $2\cdot10^5\,\Ms$ of gas.  However, should the star formation duration be longer, either because the present model is valid for more than 5$\tff$, or because the actual star formation history of molecular clumps consists of more stages than described by our model, the formation of massive bound clusters when $\eff = 0.01$ will require less extreme gas masses since both the star formation efficiency and the bound fraction will be higher.

In the second regime, when $\eff = 0.10$ (bottom panel of Fig.~\ref{fig:adb}), the FWHM of the intrinsic age distribution is driven by the half-life time of the star formation rate.  Note, however, that the FWHMs are larger than $t_{1/2}$ (see Table \ref{tbl:t1}).  This effect, especially prominent for the fast-evolving Clump~A, is a consequence of binning.  The \sfr of Clump~A drops by a factor of 2 within $50,000$ years.  Therefore, its half-life time is shorter than the age-bin width ($\Delta t = 0.1$\,Myr), and cannot be resolved in the intrinsic age distribution.  

Here, we note that the age-bin size adopted by \citet{kud12} and in this contribution (0.1\,Myr) is smaller than the uncertainties typically affecting derived ages of pre-main-sequence stars.  \citet{kud12} estimate an error of $0.25$\,Myr in their study of Wd~1 (see below), and \citet{pre12} quotes a Gaussian error of 0.5\,Myr when stellar variability, photometry errors and extinction uncertainties are taken into account (see Section \ref{sec:acc_dec}).  In strict logic, we should therefore have adopted an age-bin size of that order of magnitude.  However, we think it a valuable exercise to show that, even if stellar age errors were negligible (e.g., smaller than 0.1\,Myr), a proper reading of the half-life time of the \sfr from the binned age distribution may already be hindered when it comes to gas with high density/short free-fall times.  In the hypothetical absence of stellar age uncertainties, the age-bin size will be chosen so as to limit the Poisson noise in the star age distribution and, therefore, will depend on the size of the stellar sample and the age distribution extent.  Should the age bin be larger than 0.1\,Myr, the fast evolution of high-density gas will be concealed even more than presented here.

To model the impact of stellar age uncertainties, we assume a Gaussian error of $\sigma = 0.25$\,Myr \citep[as estimated for Wd~1 by][]{kud12} and obtain the `measured' star age distributions (solid blue lines with asterisks in Figs~\ref{fig:ad} and \ref{fig:adb}).  The measured age spreads, $FWHM_{msd}$, are given in Figs~\ref{fig:ad} and \ref{fig:adb}.  Unsurprisingly they can be significantly larger than their intrinsic counterparts.  

When $\eff = 0.10$ (Fig.~\ref{fig:adb}), the steady decrease of the \sfr leads to a measured age distribution of Clump~C more prominent on its young side than on its old side.  As for Clump~A, the asymmetry is negligible because its intrinsic age distribution is narrower than the adopted error of $\sigma =0.25$\,Myr. 
\citet{kud12} also find an asymmetry on the young side of the stellar age distribution of the cluster Wd~1.  They interpret it as resulting from the presence of binaries (their fig.~3).  Figure \ref{fig:adb} suggests that an asymmetric age distribution could also be the imprint of the \sfr decrease driven by gas depletion.

\begin{figure}
\begin{center}
\epsscale{1.0}  \plotone{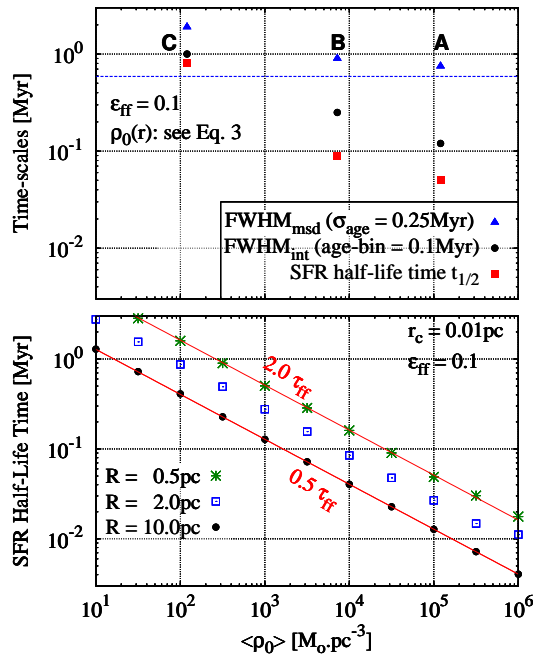}
\caption{{\it Top panel:} Half-life time of the star formation rate ($t_{1/2}$), intrinsic ($FWHM_{int}$) and measured ($FWHM_{msd}$) FWHMs of the star age distribution as a function of clump mean density ($\langle\rho_0\rangle$) when $\eff = 0.10$.  The dashed horizontal (blue) line indicates the FWHM of a Gaussian with a standard deviation of 0.25Myr, which is the assumed linear error on star ages in our simulations.  {\it Bottom panel:} Symbols show $t_{1/2}$ in dependence of $\langle\rho_0\rangle$ for three clump radii.  Lines represent half and twice the clump free-fall time. 
}
\label{fig:sum}
\end{center} 
\end{figure}

The top panel of Fig.~\ref{fig:sum} summarizes the half-life time ($t_{1/2}$), intrinsic ($FWHM_{int}$) and measured ($FWHM_{msd}$) FWHMs found for the second regime, that is, when $\eff = 0.1$.  Here we included an additional model clump, with a density between that of Clumps~A and C (Clump~B: $M_0=2.5\cdot 10^5\Ms$ and $R=2pc$).
The dependence of $t_{1/2}$ and $FWHM_{int}$ on the initial mean density of the gas, $\langle\rho_0\rangle$, is clearly highlighted.  One could argue here that the model is degenerate in the sense that narrow FWHMs can be obtained from either high-density gas, or short star-formation duration in low-density gas.  This is not the case.  For low-density clumps to give rise to $FWHM_{int}\simeq0.2$Myr, comparable to that of Clumps~A and B, star formation must be halted on a similar time-scale.  At $t\simeq 0.2Myr$, the Clump~C stellar mass is 250$\Ms$.  The star formation efficiency, $250/4000 \simeq 0.06$, gives a bound fraction of $\simeq 0.20$ \citep[fig.~15 in][]{par13}.  The star cluster mass when gas expulsion  effects are over is therefore $\simeq 50\Ms$ (i.e. $0.20\cdot0.06\cdot4000\Ms$).  Such a low-mass cluster would be swiftly disrupted by the Galactic tidal field.  It is therefore highly unlikely to observe clusters characterized by narrow age spreads {\it and} formed in low-density environments.

The bottom panel shows the relation between $t_{1/2}$ and $\langle\rho_0\rangle$.  It is obtained by setting Eq.~\ref{eq:sfrcpl} equal to $\frac{1}{2}SFR(t=0)$ and solving for $t$.  The clump radii considered are quoted in the key, although note that the parameter of relevance here is the ratio $R/r_c$.  Also shown are half and twice the free-fall time at the gas initial mean density (see Eq.~\ref{eq:tff}).  One can see that, as suggested in Section~2.2, the half-life time of the \sfr is of the order the clump free-fall time when $\epsilon_{ff}=0.1$.  

Using Eq.~\ref{eq:sfet} derived in Section \ref{ssec:mod}, we can estimate the global \sfe achieved by the clump after one half-life time (equivalent to $\simeq \tff$ when $\epsilon_{ff}=0.1$).  For clumps B and C, $\chi = r_c/R = 0.005$, $\xi = 0.1/\sqrt{3}$ and $SFE = 0.15$ at a time $t  = \tff \simeq t_{1/2}$ (the numerical solution predicts $SFE = 0.16$).  At first glance, one may be surprised that after one free-fall time, the global \sfe is higher than $\eff = 0.1$.  This effect is a direct consequence of the gas mass concentration inside the clump.  At $t=\tff$, the (three-dimensional) half-mass radius of the forming cluster is $r_{hm} = 0.45$\,pc.  This shows that the bulk of star formation takes place in the clump inner regions where the free-fall time is shorter than the free-fall time $\tff$ averaged over the whole clump.  This effect was already pointed out by \citet{tan06} and is also discussed by \citet{par14}.  For Clump~A, Eq.~\ref{eq:sfet} predicts a slightly lower global star formation efficiency, i.e., $SFE(t=\tff) \simeq 0.11$ ($SFE = 0.12$ in the numerical solution).  This is because the central core of the initial gas mass distribution occupies a greater fraction of the clump radius ($\chi = r_c/R = 0.02$), which decreases the gas mass central concentration.  After five free-fall times, the global \sfe achieved by Clumps~B and C is $SFE \simeq 0.42$ ($SFE = 0.46$ in the numerical solution).  This number reflects the decreasing star formation rate over five free-fall times of star formation activity.  These efficiencies are slightly higher than those observed, which range from a few per cent to 0.33 \citep{lad03, hig09}.  This may indicate that either the \sfe per free-fall time is slightly lower than $\eff = 0.1$, or that the duration of star formation is shorter than five free-fall times.  Note, however, that star formation efficiency estimates for embedded clusters are mostly available for the Solar Neighborhood.  In addition, we remind that not only do measured star formation efficiencies depend on time, they also depend on the spatial scale on which they are measured, with smaller scales (i.e. more limited to the cluster inner regions) leading to higher efficiencies \citep[see fig.~7 in][]{par13}.

\section{Observations}
\label{sec:obs}

\begin{table}[t]  
\begin{center}
\caption{Mean volume densities and age ranges, or FWHMs, for a sample of star-forming regions and young star clusters.  The data are detailed in Section~\ref{sec:obs} and plotted in Fig.~\ref{fig:litdat}}  
\label{tbl:t2}
\begin{tabular}{lcccc}
\tableline\tableline
                                  &  $\langle\rho\rangle$              &  Age range         &     FWHM    \\ 
                                  &  [$\Ms\cdot pc^{-3}$]  &   [Myr]            &     [Myr]                    \\ \tableline
LH95                              &    0.47               &      --            &     3.6            \\
W3 Main (stars only)              &    37                 &      3.             &     --            \\
W3 Main (stars + gas)             &    83                 &      3.             &     --            \\
ONC (stars only)                  &    56                 &      --            &     3.0             \\
ONC (stars + gas)                 &   140                 &      --            &     3.0              \\
Wd~1 ($r\leq0.2pc$)               &   $4 \cdot 10^4$     &      --            &     0.3              \\
Wd~1 ($r\leq2.9pc$)               &    $10.^3$            &      1.            &     --                \\
NGC3603 YC ($r\leq0.5pc$)         &    $17.6 \cdot 10^3$ &       --            &     0.1             \\
NGC3603 YC ($r\leq2.0pc$)         &   550                 &      --            &     8.0             \\
Chamelon II                       &   18                  &      --            &     4.0              \\
Lupus                             &   7.5                 &      --            &     5.0             \\
Ophiuchus                         &   18                  &      --            &     2.0             \\
IC348                             &   50                  &      3.5           &     --              \\ \tableline
\end{tabular}
\end{center}
\end{table}

\begin{figure}
\begin{center}
\epsscale{1.0}  \plotone{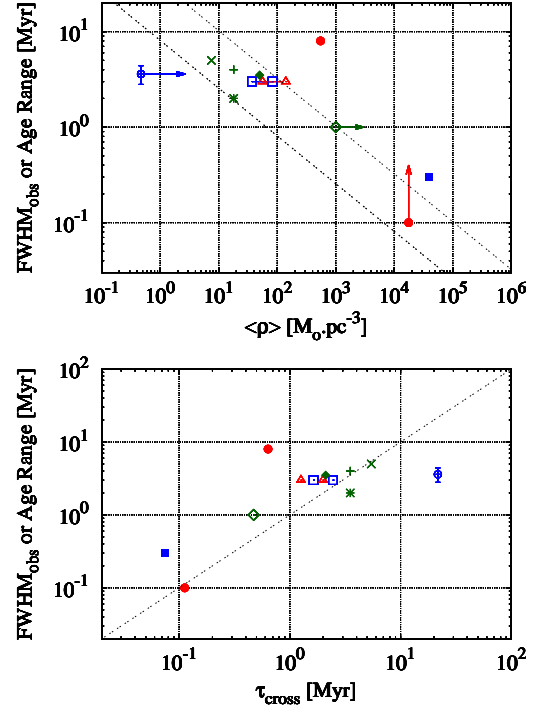}
\caption{{\it Top panel}: Age spreads as a function of mean volume density for a sample of observed star-forming regions, embedded clusters and young star clusters.  The age spread is either the measured age range or, when available, the FWHM of the linear star age distribution.  The data  are compiled in Table \ref{tbl:t2}, with full details given in Section \ref{sec:obs}.  From left to right: (blue) open circle with error bars and rightwards arrow: LH95; (green) cross, plus-sign and asterisk: Lupus, Chamaleon II and Ophiuchus molecular clouds, respectively; pair of (blue) open squares: W3 Main; pair of (red) open triangles: ONC; filled (green) diamond: IC348; filled (red) circle: NGC3603~YC (large aperture), open (green) diamond with rightwards arrow: Wd~1 (large aperture); filled (red) circle with upwards arrow: NGC3603~YC (small aperture); filled (blue) square: Wd~1 (small aperture).  The thin dashed lines indicate the free-fall time and four times the free-fall time as a function of mean volume density.  {\it Bottom panel}: Age spreads as a function of the crossing-time for the same sample of star-forming regions and clusters.  The thin dashed line indicates the locus of points where the age spread equates the crossing-time.  }
\label{fig:litdat}
\end{center} 
\end{figure}

In this section, we assess whether the currently available observations support a star formation scenario where higher-density clumps lead to narrower star age distributions.  To that purpose, we have surveyed the literature and gathered densities and age distribution widths for star clusters and star-forming regions.  They are presented in Table \ref{tbl:t2} and in the top panel of Fig.~\ref{fig:litdat}.  Depending on the available data, the width of the star age distribution is either its FWHM, or an age range estimate. 
 
When comparing such data to model predictions, there are two caveats to keep in mind.  First, our model refers to the total mean density $\langle\rho_0\rangle$ of cluster-forming clumps (i.e., gas and newly-formed stars), while the density given in Table \ref{tbl:t2} is sometimes the stellar density alone (e.g., NGC3603~YC, Wd~1).  Due to residual star-forming gas expulsion and the ensuing star loss and cluster spatial expansion, the current density of a cluster is a lower limit to the density of its parent clump.  Second, star formation is not yet over in gas-embedded clusters, which prevents the estimate of the star age spread in a fully-formed cluster.  With these caveats in mind, we now describe each data point. \\

{\bf LH95} is a star-forming region in the Large Magellanic Cloud.  Its mean stellar density is $0.47\Ms \cdot pc^{-3}$ \citep[table 2 in][]{dar09}.  From the Gaussian distribution of pre-main sequence star ages, \citet{dar10b} derive a FWHM of the order of 2.8-4.4Myr.  We adopt a mean value of 3.6Myr (open circles in top panel of Fig.~\ref{fig:litdat}).  We note that LH95 consists of three subclusters, each with a mean stellar density higher than that of LH95 as a whole \citep[fig.~4 and table 2 in][]{dar09}.  The range covered by the subcluster densities is depicted as the blue horizontal arrow.  We note that including the residual molecular gas would move the tip of the arrow further to the right. \\

{\bf W3~Main:}  This Galactic star-forming region hosts pre-main sequence stars, stars and HII regions in various evolutionary stages.  One of its massive stars is consistent with an isochrone of age 2-3Myr and its HII regions span the range from very young hypercompact (few $10^3$\,yr old) to old and diffuse ones (few $10^6$\,yr old).  \citet{bik12} therefore estimate the age range in the embedded cluster W3~Main to be at least 3Myr.  
Using their estimates for the embedded-cluster radius (3pc), the stellar mass (4000$\Ms$) and the star-plus-gas mass (9000$\Ms$), we derive stellar and total volume densities of $37$ and $83\Ms \cdot pc^{-3}$.  The corresponding data are shown as the blue open squares in Fig.~\ref{fig:litdat}.      \\

{\bf The Orion Nebula Cluster (ONC):}  \citet{hil98} estimate its stellar and total masses within a radius of 2pc to be 1800$\Ms$ and 4500$\Ms$, respectively.  The corresponding volume densities are 56 and 140$\Ms \cdot pc^{-3}$.  We obtain the FWHM of the linear distribution of pre-main sequence star ages from fig.~11 in \citet{dib13}, which is based on the data of \citet{dar10b}.  Assuming a symmetric age distribution gives a FWHM of $\simeq 3$\,Myr.  These results are depicted as the open red triangles in Fig.~\ref{fig:litdat}.  \\

{\bf Wd~1} is a young compact massive cluster of the Galactic disk.  Its age, mass, and half-light radius are estimated to be, respectively, 5\,Myr \citep{kud12}, $10^5\Ms$ and 0.86\,pc \citep{men09}.  \citet{kud12} obtain the age distribution of the stars of the cluster central region (stars with masses between 0.5 and 11.5\,$\Ms$ within a field-of-view 0.2\,pc in radius). They find $FWHM_{obs} = 0.4$Myr.  Subtracting in quadrature their photometric error estimate (0.25Myr) leads to $FWHM_{int} = 0.3$Myr.  We still need the mean stellar density within 0.2pc of the cluster center.  For a Plummer model with the above-quoted mass and half-light radius, the stellar mass enclosed within 0.2pc from the cluster center is 1200$\Ms$ \citep{hh03}.  The corresponding mean volume density is $3.8 \cdot 10^4\Ms \cdot pc^{-3}$.  This point is shown as the blue filled square in Fig.~\ref{fig:litdat}. \\
\citet{neg10} consider a field-of-view an order of magnitude larger than that of \citet{kud12} (radius: 2.9\,pc).  They find that the star age range is very unlikely to be $>1$ Myr.  Assuming again a Plummer model, the mean volume density within 2.9\,pc from the cluster center is $\simeq 10^3\Ms \cdot pc^{-3}$.  The corresponding result is shown as the green open diamond in Fig.~\ref{fig:litdat}.  We note that the cluster has likely expanded and lost mass (gas and stars) as a result of residual star-forming gas expulsion \citep{par12}.  This correction would move the point to higher densities as indicated by the horizontal green arrow (the arrow size is arbitrary).  As for the central region observed by \citet{kud12}, its \sfe was probably higher than in the cluster outskirts \citep{par13}, making it less prone to post-gas-expulsion expansion and better able to retain the initial density. \\

{\bf \3 YC}\footnote{We refer this young cluster as \3 YC to make it distinct from the giant HII region NGC~3603 at the center of which it is located} is also a starburst cluster of the Galactic disk, with a dynamical mass estimate of 17,600$\Ms$ and a (uncertain) half-mass radius of 0.5pc \citep{roc10}.  The study of its age distribution has yielded very contrasting results so far.  \citet{kud12} infer a FWHM of 0.1Myr from their linear star age distribution.  The sample includes main sequence, pre-main sequence stars and transition region stars with masses between 0.8 to 6.5\,$\Ms$.  
Based on proper-motion cluster membership of fainter main-sequence
stars and the apparent width of the region where pre-main-sequence
stars join the main sequence, \citet{pan13} find support for a
modest age spread in the cluster, or for earlier star formation
episodes in that region.
Based on pre-main sequence stars, \citet{bec10} estimate that star formation in and around the cluster has been ongoing for 10-20Myr.  In what follows, we assume that the linear FWHM of \citet{bec10} is 8\,Myr (by inspecting their log-log star age distribution; see their fig.~8).  
How can two FWHM estimates \citep{kud12,bec10} of the same cluster differ by almost two orders of magnitude?  Part of the difference resides in both studies scrutinizing different regions of the cluster.  \citet{kud12} study the region within 0.5pc from the cluster center, where the mean density is $1.8 \cdot 10^4\Ms \cdot pc^{-3}$ (using a Plummer model, the half-mass radius and total mass quoted above).  In contrast, \citet{bec10} consider a wider field-of-view, with a radius of about 2pc and an enclosed mean density of $\simeq 550 \Ms \cdot pc^{-3}$.  Both results are shown as solid red circles in Fig.~\ref{fig:litdat}.  Besides the difference in the field-of-view size, we note that the FWHM inferred by \citet{kud12} is surprisingly small, being even lower than the photometric error they derive for Wd~1.  Given that they do not discuss the photometric error contribution to the age spread of \3 YC, we consider their value of 0.1Myr a lower limit which we indicate by an upward arrow in the top panel of Fig.~\ref{fig:litdat}.  We also note that \citet{kud12} include main sequence stars in their sample while \citet{bec10} consider pre-main sequence stars only.  We further discuss the difference between both studies later in this section.     \\

{\bf IC348} is an embedded cluster of the Perseus molecular cloud.  It contains about 400 members within a radius of $\simeq 1$pc \citep{luh03}.  Adopting a mean stellar mass of 0.5$\Ms$, the mean stellar density of IC348 is $\simeq 50 \Ms \cdot pc^{-3}$.  Star formation started 3.5Myr ago \citep{mun03} and is continuing today \citep{hat05}.  We therefore adopt an age range of 3.5Myr when plotting the corresponding point in Fig.~\ref{fig:litdat} (filled green diamond).  Note that including the gas mass would move the point rightwards. \\

{\bf Cha II, Lupus, and Ophiuchus} are star-forming molecular clouds of the Solar Neighborhood. 
We infer their mean volume density from the masses and surface areas given in table~1 of \citet{eva09}, assuming spherical symmetry.  The FWHM of their linear star age distribution are estimated by inspecting figs~2, 1 and 3 in \citet{pal99}.  Here, we assume that each age distribution is symmetric about its peak, which may or may not be true. 
The mean volume densities and FWHMs are 18$\Ms \cdot pc^{-3}$ and 4\,Myr (Cha~II), 7.5$\Ms \cdot pc^{-3}$ and 5\,Myr (Lupus), 18 $\Ms \cdot pc^{-3}$ and 2\,Myr (Ophiuchus).  They are shown as the plus-sign, cross and asterisk in Fig.~\ref{fig:litdat}.  Here, we repeat the caveat already made in Section \ref{sec:intro}, that is, our model applies to molecular clumps inside molecular clouds and not to molecular clouds as a whole.  The YSO age distribution for a whole cloud is necessarily widened  by the age differences between its constituent molecular clumps (i.e., not all clumps started to form stars at the same time).  Ideally, one would use the mean volume density and YSO age distribution of star-clustering region {\it individually}.  This would move the symbols in Fig.~\ref{fig:litdat} towards higher densities and smaller age spreads.             \\
     
The data collected in the top panel of Fig.~\ref{fig:litdat} suggest that denser clusters have narrower star age distributions.  However, this conclusion is mostly driven by the central regions of the starburst clusters Wd~1 and \3 YC.  It is highly desirable to add more data points to the high-density regime ($\langle\rho\rangle > 10^4\Ms \cdot pc^{-3}$) to see whether the highlighted trend is confirmed or not.  Whether the correlation stands within the low-density regime ($\langle\rho\rangle < 10^2\Ms \cdot pc^{-3}$) cannot be established.  The data points occupy a fairly limited region of the parameter space so that any trend is likely concealed by the uncertainties affecting the age range/spread and volume density estimates.  Clusters with intermediate densities (i.e. $10^2 < \langle\rho\rangle < 10^4\Ms \cdot pc^{-3}$) would also be helpful.  In that respect, NGC~1333 in the Perseus molecular cloud is an interesting target.  With a gas mass of $10^3\Ms$ enclosed within a radius of 0.5pc \citep{lad03}, its mean density is $2000 \Ms \cdot pc^{-3}$.  \citet{wil04} infer a median cluster age of less than 1Myr.  To constrain the age spread of this forming cluster would allow to populate the diagram in the intermediate-density regime.   

The bottom panel of Fig.~\ref{fig:litdat} presents the age spread as a function of the crossing-time for the same sample of star-forming regions and clusters.  We define the crossing-time as $\tau_{cross} = 1/\sqrt{G\rho}$.  Therefore, $\tau_{cross} \simeq 2\tff$.  As already found by \citet{elm00}, the age spread is of order 1-2\,$\tau_{cross}$, or $2$-$4\tff$.  \\

We now come back to the age spread of \3 YC, and to the very different results obtained by \citet{bec10} and \citet{kud12}. 
The data point for \3 YC by \citet{bec10} stands out of the density-vs-age-spread correlation.  Their estimated age spread is about 2 orders of magnitude larger than found by \citet{kud12} (see red circles in Fig.~\ref{fig:litdat}).  As noted above, both studies differ by the size of their field-of-view (radii of 2.9pc and 0.5pc, respectively).  \citet{bec10} therefore include lower-density regions where star formation proceeded on a slower time-scale, naturally yielding a wider age distribution.  A larger aperture also encompasses more cluster stars located along the line-of-sight and more distant from the cluster center than 2.9\,pc.  This broadens further the observed star age distribution.  This effect is akin to the central (denser) regions of star clusters returning faster to virial equilibrium than the cluster as a whole, as highlighted by \citet{par12}.  (A larger aperture also increases the probability of contaminating the sample with stars not belonging to the cluster).  We therefore stress that stellar age spreads are aperture-dependent.  We call for a consistent comparison of literature results in terms of aperture size and nature of the stellar sample (main sequence vs. pre-main sequence stars) under scrutiny.  

The very large age spread found by \citet{bec10} might also be driven by pre-main sequence stars being erroneously assigned old ages.  \citet{man13} have recently reconsidered two major outliers of the stellar age distribution of the ONC.  Their estimated isochronal age is $\gtrsim 30$Myr and they show strong accretion activity (as indicated by their H$\alpha$ emission excess).  For these outliers, \citet{man13} have obtained new age estimates in agreement with the age of the bulk of the ONC population, i.e. $\simeq 2.5$\,Myr.  The cause for this age shift was an incorrect estimate of either the spectral type or the visual extinction.  \citet{man13} suggest that these issues affect mostly young star-forming regions where effects such as extinction and the high accretion rate from the inner protostellar disc onto the YSO surface are significant \citep[see also][]{pan13}.  \\

\section{Accelerated vs. decelerated star formation}
\label{sec:acc_dec}

Our model predicts a decrease with time of the star formation rate.  The higher the clump volume density and/or the star formation efficiency per free-fall time, the faster the \sfr decreases.  For low-density regions with $\eff = 0.01$, the \sfr is almost constant in time (e.g., Clump~C, Fig.~\ref{fig:sfhb}).  
Let us assume for now that the mass of the newly formed stars is independent of time, that is, the stellar {\it mass} formed per time unit and the {\it number} of stars formed per time-unit evolve in a similar way.
Our results are then supported by the observations of \citet{hat05}.  In a submillimetre continuum survey of two embedded clusters of the Perseus molecular cloud, IC348 and NGC~1333, they find evidence for a steady or reduced (number) star formation rate in the last 0.5\,Myr, not an increasing one.  Note, however, that their results depend on their assumed duration of the Class0/I phase.  Indications of a star formation activity declining with time have also been obtained for the star-forming region Chamaeleon~I by \citet{luh07} and \citet{bel11}.

\begin{figure}
\begin{center}
\epsscale{1.0}  \plotone{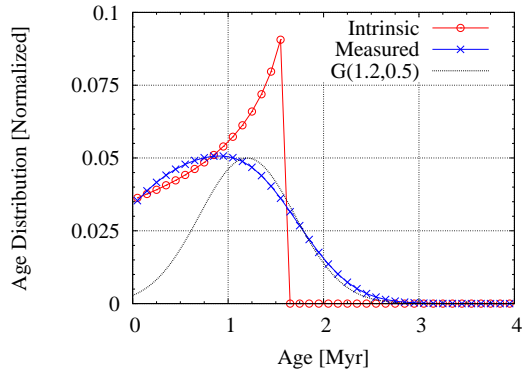}
\caption{Intrinsic and measured star age distributions of a clump where star formation has been ongoing for 1.5\,Myr.  The clump mass and radius are $4000\,\Ms$ and $2\,pc$, and the \sfe per free-fall time is 0.1.  Star age estimates are affected by a Gaussian error of 0.5Myr (standard deviation), represented by the dotted black line. }
\label{fig:acc}
\end{center} 
\end{figure}

Our concept of a decreasing \sfr contrasts with the scenario of \citet{pal00} in which the (number) star formation rate accelerates with time.  We now discuss this aspect.
It remains uncertain whether the increase with time of the star formation rate observed by \citet{pal00} for a number of embedded clusters and star-forming molecular clouds of the solar neighborhood is genuine or not.  \citet{pre12} emphasizes that the error affecting isochronal age estimates of pre-main sequence stars can by itself carve an apparent decrease towards old age in the linear star age distribution.  We illustrate his point with our model Clump~C in Fig.~\ref{fig:acc}.  We assume that star formation has been ongoing for 1.5Myr with a \sfe per free-fall time of 0.10 and we convolve the intrinsic age distribution with a Gaussian error of 0.5Myr (standard deviation).  This error is that estimated by \citet{pre12} for a population of pre-main sequence stars with masses between 0.1 and 7\,$\Ms$.  It includes effects such as stellar variability, photometry errors and extinction uncertainties.  While the high \sfe per free-fall time results in a significant decrease in the number of stars per age bin towards young ages (red curve in Fig.~\ref{fig:acc}), the age distribution convolved with the Gaussian error shows an {\it apparently} increasing star formation rate on its old side (see blue curve).  The comparison between the right sides of the measured age distribution and of the Gaussian function (depicted as the dotted black line) shows this apparently accelerating star formation rate to be the signature of the stellar age error, as warned by \citet{pre12}.  

\citet{bec10} suggest yet another reason as to why the observed (number) \sfr increase may be apparent only.  That is, many of the old pre-main sequence stars may have migrated out of the field-of-view, which artificially decreases the number of stars at old ages.  This suggestion is supported by their study of \3 YC, in which they find a correlation between the estimated ages of pre-main sequence stars and their location in the cluster.  Their fig.~7 shows young pre-main sequence stars (isochronal age $<$10Myr) to be more centrally concentrated than their older counterparts (isochronal age $>$ 10Myr).  This may be a hint that the old objects are migrating away from the cluster center and observed field-of-view, thereby causing an apparent \sfr increase \citep[but see the caveat about age estimates of old pre-main sequence stars by][]{man13}, or that they
are indeed unrelated to the cluster and the product of an earlier
star formation episode \citep[see, e.g., ][]{pan13}.

These two examples -- errors affecting the age estimates of pre-main sequence stars and stellar motions removing them from the observed field-of-view -- demonstrate the difficulty to infer the actual evolution of the star formation rate, even in nearby star-forming regions.  Current observational data hardly allow to disentangle between an intrinsically increasing \sfr and an intrinsically decreasing one. 

Our model predicts a (mass) \sfr decreasing with time, starting from an isolated star-less clump.  Here we re-iterate that the steady decrease of the \sfr stems from the model assumptions: the sudden onset of star formation in a molecular clump whose outer radius remains constant in time.  Should star formation already start as the clump contracts, there will be an initially increasing star formation rate, although that does not imply that most stars form during that phase.
A (mass) \sfr genuinely increasing with time is the expected star formation history of a contracting molecular clump, since the density increase of the collapsing gas shortens the clump free-fall time.  This scenario naturally leads to the outer regions of the forming star-cluster to be populated by the first stars to form, hence the oldest ones.  
The actual evolution of a star-forming molecular clump may combine a contraction (\sfr increase) followed by a near-equilibrium evolution (\sfr decrease) once stellar feedback has halted the contraction \citep{kru06, nak07}.

Not only is the issue of the star formation rate evolution important in its own right, it also bears consequences for the estimate of the Class~I phase duration.  \citet{eva09} estimate it by taking the ratio of number counts in Class~I and Class~II stages, and multiplying by the assumed lifetime of the Class~II phase.  By so doing, they assume that star formation has been proceeding at a constant rate over a time-interval longer than the added durations of the Class~I and Class~II phases.  In our model, this hypothesis holds in the low-density star-forming regions of the solar neighborhood, provided that the \sfe per free-fall time is not higher than a few per cents.   If the star formation rate decreases (as expected in our model when $\eff = 0.1$), the Class~I lifetime estimate should be extended.  In contrast, if the star formation rate increases with time, as suggested by \citet{pal00}, then the Class~I phase duration is to be shortened.  \\

The comparison between the mass \sfr predicted by a model and the number star formation rate retrieved from the observations may become even more intricate than presented above if the mass of stars depend on the time of their formation.  Should that be the case, the star formation rates per mass and number evolve differently.  For instance, if the high-mass stars form first, the mass \sfr declining with time predicted by our model will lead to a smoother decline of the number \sfr or, possibly, to an accelerating number star formation rate.  Additionally, the detection of the stars with the lowest luminosity is hindered by the presence of the star-forming gas.  Therefore, the stellar sample may be incomplete for the low-mass and/or old pre-main sequence stars, making the model-observations comparison even less straightforward.  To discuss these two points, however, is well beyond the scope of the present paper.

\section{Discussion and Conclusions}
\label{sec:conclu}

We have modeled the stellar age distributions of young clusters based on the density-dependent star formation model of \citet{par13}.  In this class of models, the star formation efficiency increases more rapidly in higher volume-density gas than in low volume-density gas.  Our model starts with starless molecular clumps in equilibrium, and we assume that a static configuration is retained throughout the star formation episode.  For this reason, we limit the duration of star formation to five initial free-fall times of the clump.  
We also assume that the clump neither gains nor loses mass.  Therefore, the \sfr is steadily decreasing with time since, as star-formation depletes the clump gas, the \sfe per free-fall time is applied to an ever lower amount of gas on an ever longer time-scale.    

Our results are equations describing the star formation rate evolution of a clump with given mass and radius   (Eq.~\ref{eq:sfr2}).  We also provide analytical equations obtained for the specific case of an isothermal sphere of gas (Eqs.~\ref{eq:sfrcpl} and \ref{eq:cround}). 

For each of our model clumps, we have presented the star formation history and the linear distributions of the star ages if the newly-formed cluster is observed 5\,Myr after the onset of star formation.  Regarding the star age distribution, we have considered the {\it intrinsic} age distribution, where star ages are binned with a bin size of 0.1\,Myr, and the {\it measured} age distribution, which accounts for the errors affecting individual star ages.  That is, the measured age distribution is the intrinsic one convolved with a Gaussian function.  We have successively discussed the star formation history of the clump in relation to the clump properties (mainly mean density and \sfe per free-fall time; see Figs.~\ref{fig:princ} and \ref{fig:princb}) and how the star formation history translates into the cluster star age distribution (Figs.~\ref{fig:ad} and \ref{fig:adb}).  

In our study of the clump star formation history, we have considered two regimes, depending on whether the \sfr decreases by a factor of 2 during star formation or not.
If the evolution is rapid enough for it to decrease by a factor of 2, one can define the half-life time of the clump star formation rate, namely, the time needed for the \sfr to decrease to half its initial value.
This half-life time is a key-driver of the FWHM of the cluster intrinsic age distribution. 
If the \sfr has decreased by less than a factor of two at the end of the star-formation episode, the half-life time of the \sfr cannot be defined, and the FWHM of the intrinsic age distribution is driven by the duration of star formation.  We have embodied these two regimes with two distinct star formation efficiencies per free-fall time, $\eff = 0.1$ and $\eff = 0.01$.

When $\eff = 0.1$ (rapid evolution), the half-life time of the \sfr is of the order of the free-fall time at the clump mean volume density.  (Note that this result is valid for an isothermal sphere of gas.  Shallower density profiles lead to longer half-life times.)  Therefore, the denser the cluster-parent clump, the shorter the half-life time of its star formation rate and the shorter the stellar age spread of the resulting cluster.  

Caution should be taken, however, as the FWHM of the age distribution does not depend only on the half-life time of the star formation rate.  Firstly, star-age binning may prevent a short half-life time from being resolved in the intrinsic age distribution.  This is the case encountered for our densest model clump (Model~A).
Secondly, uncertainties in the age of individual stars inflate the measured age spread.  
In case of a very dense -- fastly-evolving -- clump, the measured age spread embodies mostly the star age uncertainty and loses memory of the short \sfr half-life time.    
Therefore, these two effects -- large age-bin size and age uncertainties -- weaken the contrast between the measured age spread for high- and low-density clumps (see Fig.~\ref{fig:sum}, top panel).  

The predicted effect that high-density clumps give rise to narrower age spreads seems to have been observed, with the FWHM of the stellar age distribution in the central regions of the high-density clusters NGC3603~YC and Wd~1 found to be an order of magnitude smaller than for the Orion Nebula Cluster \citep[see][]{reg11, kud12}. 
For the central regions of the young starburst cluster Wd~1, \citet{kud12} derived a FWHM of 0.4Myr.  They estimated a photometric error contribution of 0.25 Myr.  Assuming that the observed FWHM is the quadratic sum of the age error and of the intrinsic FWHM, we derive $FWHM_{int} \simeq 0.3$Myr.  This value agrees well with what we found for Clump~B (see top panel of Fig.~\ref{fig:sum}).  We therefore suggest that the central regions of Wd~1 formed out of gas with a density of the order of $10^4\,\Ms \cdot pc^{-3}$ if $\epsilon_{ff}=0.10$.  

Matching a cluster age spread to the density of its parent clump can be done by using the bottom panel of Fig.~\ref{fig:sum}.  It provides the relation between the half-life time and the mean volume density of molecular clumps with an $r^{-2}$-density profile when $\eff = 0.1$. Observers can build on it to predict the relation between the clump mean density and the FWHM of the stellar age distribution they would measure for their own age-bin size and uncertainties in stellar ages. It can also be rescaled to another $\epsilon_{ff}$-value: lower efficiencies lead to higher gas densities for given FWHMs. 
If the \sfe per free-fall time is lower than 0.1, the evolution is slower.  This implies that to achieve a given age spread $FWHM_{int}$, the free-fall time of the gas must be shorter (to compensate for the slower $\eff$) and, therefore, the gas density must be higher.  As an example, if an intrinsic age spread $FWHM_{int}$ is achieved by a clump of mean density $\langle\rho_0\rangle$ with $\epsilon_{ff}=0.1$, the same age spread is achieved by a clump of mean density $4\langle\rho_0\rangle$ if $\epsilon_{ff}=0.05$.

But what if the \sfr decreases so slowly that its half-life time cannot be defined?  If the \sfe per free-fall time is as low as $\eff = 0.01$, it takes about 10 clump free-fall times for the \sfr to decrease to half its initial value (again assuming an isothermal density profile).  Therefore, if the star formation duration is shorter than that, the half-life time cannot be defined, and the FWHM of the intrinsic star age distribution is driven by the duration of the star formation episode $\Delta t_{SF}$ (e.g. bottom panel of Fig.~\ref{fig:ad}).
To estimate $\Delta t_{SF}$ requires a detailed understanding of how long it takes for stellar feedback processes to terminate star formation inside the clump, which is beyond the scope of this paper.  Yet, we have shown that building on the invariance of the shape of the {\it young} cluster mass function can help us reach some conclusions.    
That the shape of the cluster mass function does not change with time at young ages points to a mass-invariant \sfe and, therefore, to a duration of star formation equal to a given number of free-fall times for all cluster-forming clumps.  Therefore, in this case too, we reach the conclusion that the $FWHM_{int}$ scales with the clump free-fall time $\tau_{ff}$, i.e. $FWHM_{int} \simeq \Delta t_{SF} = k\cdot \tff$.  Again, denser clumps give rise to narrower star age distributions.  The normalisation of the $\langle\rho_0\rangle$-$FWHM_{int}$ relation depends on how many free-fall times star formation lasts.

We have also searched the literature for the age spreads and densities of observed star-forming regions.  We find that current data support a scenario where the characteristic time-scale of star formation is of order $1$-$4\tff$, which is reminiscent of the earlier study by \citet{elm00} (see Fig.~\ref{fig:litdat}).  Note that the $y$-axis of Fig.~\ref{fig:litdat} often refers to the observed FWHM of the stellar age distribution.  In our model, it reflects the half-life time $t_{1/2}$ of the star formation rate.  As such, the FWHM is not representative of the duration of star formation which can be significantly longer.

The key conclusion of our work is that the observation in young clusters of stellar age spreads spanning 1-to-2 orders of magnitude \citep{jef11, reg11, kud12} does not require different cluster-formation mechanisms \citep{dib13}.  This may simply reflect the wide range of volume densities among cluster-forming clumps.    
Our work also suggests that the intrinsic star age spread in clusters may constitute a promising probe into the mean volume density of their gaseous progenitors, provided that the \sfe per free-fall time is well-constrained.



\acknowledgments
GP acknowledges support from the Olympia-Morata Program of Heidelberg University.  This work was also supported by the Sonderforschungsbereich SFB 881 "The Milky Way System" (subproject B5) of the German Research Foundation (DFG).  We thank Nicola Da Rio and Rob Jeffries for interesting discussions, and the referee for a thorough report.







\end{document}